\documentstyle[11pt,aaspp4,flushrt]{article}
\newcommand\beq{\begin{equation}}
\newcommand\eeq{\end{equation}}

\begin{document}

\title{Jets in GRBs: Tests and Predictions for the Structured Jet Model}

\author{Rosalba Perna\altaffilmark{1,2,3}, Re'em Sari\altaffilmark{3} 
and Dale Frail\altaffilmark{4}}

\altaffiltext{1}{Harvard Society of Fellows, 74 Mount Auburn Street, Cambridge, MA 02138}

\altaffiltext{2}{Harvard-Smithsonian Center for Astrophysics, 60 Garden Street,
Cambridge, MA 02138}

\altaffiltext{3}{130-33, California Institute of Technology, Pasadena, CA 91125}

\altaffiltext{4}{National Radio Astronomy Observatory, P.O. Box O, Socorro, NM, 87801}

\begin{abstract}

The two leading interpretations of achromatic breaks that are observed
in the light curves of GRBs afterglow are (i) the manifestation of the
edge of a jet, which has a roughly uniform energy profile and a sharp
edge and (ii) a line of sight effect in jets with a variable energy
profile. The first scenario requires the inner engine to produce a jet
with a different opening angle each explosion, while the latter
requires a standard engine.  The physical structure of the jet is a
crucial factor in understanding GRB progenitors, and therefore
discriminating the two jet scenarios is particularly relevant.  In the
structured jet case, specific predictions can be made for the
distribution of observed break angles $\theta_{\rm break}$, while that
distribution is arbitrary in the first scenario. We derive the
theoretical distribution for the structured jet model. Specifically,
we predict the most common angle to be about 0.12 rad, in rough
agreement with the sample. If this agreement would hold as the sample
size increases, it would strengthen the case for the standard jet
hypothesis. We show that a prediction of this model is that the
average viewing angle is an increasing function of the survey
sensitivity, and in particular that a mission like {\em Swift} will
find the typical viewing angle to be about 0.3 rad.  The local
event rate predicted by this model is $R_{\rm GRB}(z=0)\sim 0.5 $
Gpc$^{-3}$ yr$^{-1}$.

\end{abstract}


\section{Introduction}

The degree to which gamma-ray bursts (GRBs) and their afterglows are
beamed is an important issue. A proper understanding of the geometry
of the relativistic outflow affects the total energetics of GRB
central engines and the GRB event rates, both of which are crucial
parameters for constraining possible progenitor models.  Evidence for
non-isotropic outflows is believed to come from observations of
achromatic breaks in afterglow light curves (e.g., Rhoads 1999, Sari
et al. 1999) and the detection of polarized emission (Covino et
al. 1999, Wijers et al. 1999).  We should however note that, whereas
this is a natural interpretation for the breaks, in some cases other
interpretations have been proposed or different conclusions derived
(e.g. Nicastro et al. 1999; Vrba et al. 2000).  Here we adopt the
point of view that the observed breaks are indeed manifestations of
jets.

In early theoretical papers (e.g., Rhoads 1997, Sari et al. 1999) it
was assumed for simplicity that the ejecta had to be distributed
approximately uniform across the entire opening angle in the gamma-ray
phase and that the majority of the explosive energy in the afterglow
phase must have a single bulk Lorentz factor. In this ``uniform''
model, a break occurring in the light curves at $t_{\rm break}$ can be
directly translated into a jet with an opening angle $\theta_{\rm
break}$. Using this simple framework, Frail et al.~(2001) carried out
an analysis of all known afterglows and found that there was a
distribution of jet opening angles leading to a reduction in the
gamma-ray energy from its isotropic value with relatively small
scatter. The observed distribution was shown to be heavily weighted
towards small opening angles.

Rather than positing a uniform jet, it is equally reasonable to assume
that GRB jets are structured in some fashion. In the collapsar
progenitor model (e.g. Wang, Woosley \& MacFadyen 2002) the Lorentz
factor and energy are high near the rotation axis, but decrease off
axis as the degree of entrainment increases. Salmonson (2000) has
argued for such a jet structure to explain the empirical correlation
between the GRB peak luminosity and pulse lag (Norris et al. 2000). In
this case, the distribution of observed break times in afterglow light
curves are not due to a distribution of opening angles but rather
originate from variations in the viewing angle of a structured jet
(Postnov, Prokhorov \& Lipunov 2001). In two recent papers by Rossi,
Lazzati \& Rees (2002a) and Zhang \& Meszaros (2002) it was shown that
a jet with a universal beaming configuration could reproduce the
near-constant energy result of Frail et al.~(2001) provided that the
energy per unit solid angle (and Lorentz factor) varied as the inverse
square of angular distance from the jet symmetry axis.

Discriminating between the uniform and structured jet models is
important as they yield different estimates for the true GRB event
rate and the total energy, besides leading to clues on the physical
mechanism producing the jet itself. Afterglow lighcurves have argued
to be degenerate to the structure of the jet (Rossi et al. 2002a,
Zhang \& Meszaros 2002) and therefore they cannot be used as
diagnostics.  Rossi et al. (2002b) showed that detection of
polarization can provide useful constraints.  Here we concentrate on
diagnostics based on geometrical effects.  In particular, in the
structured jet model, since the inferred opening angle is just a
geometric effect of the viewing angle, it is possible to predict the
distribution of angles and compare it to the {\it observed}
distribution of $\theta_{\rm break}$ by Frail et al. (2001) and Bloom
et al.~ (2003).  Unfortunately, a specific prediction is not possible
for the uniform model, since there is no framework for jet formation
which yields the distribution of opening angles.

In this paper, we work out the expected distribution of burst
opening angles, under the structured jet scenario and given various
assumption of the star formation rate evolution in the universe.  We
find two competing effects. First, even though randomly oriented
bursts would be rarely observed on axis, they are much brighter and
therefore can be seen to larger distances rendering small opening
angles common. Cosmological effects limit the volume at large
redshifts leading to an effective cutoff at the small opening angle.
As a result, we predict that the most common opening angle should be
about 0.12 rad. Furthermore, under the structured jet model, we
predict that more sensitive future missions like SWIFT, will find a
much larger typical angle, about 0.3 rad.

The observed data so far is still a small sample. Selection effects
are hard to quantify, especially for large and small opening
angles. In addition, the variety of instruments with different
sensitivity used to detect the bursts, makes a robust conclusion
difficult at this stage. Yet, we show that the current set of data, as
given in the most updated sample of Bloom et al. (2003), is at least
in rough agreement with the distribution we predict for the structured
jet model.  If this conclusion would hold with much larger and less
biased sample, it would be a strong support to the structured jet
model.

\section{Computation of the observed distribution of jet angles}

\subsection{Scalings for a Euclidian universe}

If all bursts were observable, then we would expect that the number
$dn(\theta)/d\theta$ of bursts with angle in the interval $d\theta$
around $\theta$ would be proportional to $\theta$. This implies that
most of the observed bursts should have a large angle, which is in
complete contradiction with observations. Zhang \& Meszaros (2002)
attributed this apparent discrepancy to the small sample size or
afterglow selection effects. However, this argument does not account
for the fact that bursts with small $\theta$ are brighter by a factor
of $\theta^{-2}$, and therefore can be seen (in an Euclidian universe)
up to a distance $\theta^{-1}$ farther, which contains a volume larger
by a factor of $\theta^{-3}$. The expected distribution in Euclidean
geometry is therefore expected to be $dn/d\theta \propto
\theta^{-2}$. Although this is closer to what observations suggest, we
will show that this now exaggerates the number of small-angle bursts
compared to a proper cosmological prediction: GRBs originate at
redshifts of order unity, and therefore suffer considerable
cosmological corrections. This is why their log $N$ - log $S$ curve
does not obey the Euclidian $S^{-3/2}$, but is shallower at low
$S$. It is for the same reason that the number of observed bursts of
low $\theta$ will not be as high as predicted by the Euclidian
$\theta^{-2}$. In the following, we work out these cosmological
effects in detail.

\subsection{Cosmological effects}

Let $R_{\rm GRB}(z)$ be the GRB rate per unit comoving volume per unit
time, then the total (i.e. over the all sky) rate of bursts with
inferred jet angle between $\theta$ and $\theta\; + \;d\theta$ is
given by
\begin{equation}
\frac{dn(\theta)}{d\theta} = \sin\theta \int_0^{z_{\rm max}(\theta)}
dz\; \frac{R_{\rm GRB}(z)}{(1+z)}\; \frac{dV(z)}{dz}\;,
\label{eq:dndt}
\end{equation}
where $z_{\rm max}(\theta)$, is the maximal redshift up to which we
can observe a burst with apparent angle $\theta$. This redshift is
found by numerically inverting the equation
\begin{equation}
F_{ph, \rm lim}=\frac{L_{ph}(\theta)}{4\pi D^2(z_{\rm max})(1+z_{\rm
max})^\alpha}\;,
\label{eq:flim}
\end{equation}
where $F_{ph, \rm lim}$ is the limiting photons flux (photons per unit
area per unit time) that is detectable by the GRB detector with
frequency range, $\nu_l<\nu<\nu_u$. $L_{ph}(\theta)$ is the photons
luminosity (photons per unit time) in the same frequency rage but in
the local frame of the burst, of a burst with an apparent angle
$\theta$. A factor of $(1+z)^{\alpha-1}$ is a spectral correction,
assuming that the GRB has a differential photon spectral index
$\alpha$, and another factor of $1+z$ takes care of time dilation.
For BATSE, $\nu_l=50$ keV and $\nu_u =300$ keV.

The normalization constant is determined by the condition
\begin{equation}
\frac{L_{\nu_1-\nu_2}}{4\pi}\,2\pi\theta^2 T=E
\end{equation} 
where $E \approx 10^{51}\;{\rm ergs}$ is the roughly constant 
energy of GRBs as inferred by (Bloom et al. 2003). 
That relation used the luminosity integrated over the
frequency range $\nu_1=20$ keV to $\nu_2=2000$ keV as calculated by
Bloom et al. (2001). For powerlaw photon spectrum, this is related to
the photon luminosity in the triggering band $\nu_l$ $\nu_h$ by
\begin{equation}
L_{ph}(\theta)={2E \over \theta^2 T h\nu_l} {\alpha-2 \over \alpha -1}
\left( \nu_l \over \nu_1 \right)^{-\alpha+2}
{1-(\nu_u/\nu_l)^{-\alpha+1} \over 1-(\nu_2/\nu_1)^{-\alpha+2}}
\end{equation}

If we take a spectrum with $\alpha \approx 1$, (which is appropriate
for the frequency range $50-300$ keV as reported by Mallozzi et al. 1996) then we
obtain,
\begin{equation}
L_P(\theta)= 1.1\times 10^{57} \;T^{-1}\theta^{-2}\;{\rm ph}\;{\rm
sec}^{-1}\;.
\label{eq:Lp0}
\end{equation}
It should be noted that $T$ is not the total duration of the burst,
but an ``effective'' duration of convenience here, that is the
duration that the burst would have if its energy output were constant
at the peak value rather than highly variable.  In the simplest
version of our model (\S 2 \& 3), we will assume a single value of $T$
for all the bursts; however we will explore (\S 4) how our results
vary when a scatter in $T$ is introduced (which is a more realistic
assumption).

The jet model with the energy profile $\propto\theta^{-2}$ also makes
detailed predictions for the observed GRB flux distribution.  This,
within our formalism, can be written as \beq
\frac{dn(S)}{dS}=\int_0^\infty dL_p \;f(L_p) \left[\frac{R_{\rm
GRB}(z)}{(1+z)}\;
\frac{dV(z)}{dz}\left|\frac{dz}{dS}(z,L_p)\right|\right]_{z=z(S,L_p)}\;
\label{eq:dnds}
\eeq
where, given Eq.~(\ref{eq:Lp0}), the luminosity function takes the form 
$f(L_p)\propto L_p^{-2}$.

In Eqs.(\ref{eq:dndt}) and (\ref{eq:dnds}), $dV(z)/dz$ is the comoving volume. 
In a flat cosmology with a cosmological constant it is given by
\beq
\frac{dV(z)}{dz} = 4\pi {D^2(z)}\frac{dD(z)}{dz}\;
\eeq
where $D(z)$ is the comoving distance, 
\beq
D(z) = {c \over H_0} \int_0^z {dz' \over \sqrt{(1+\Omega_mz)(1+z')^2-\Omega_\Lambda(2z'+z'^2)} } \;,
\eeq
We assume a cosmological model with $\Omega_m=0.3$,
$\Omega_\Lambda=0.7$ and $H_0=71$ km s$^{-1}$ Mpc$^{-1}$. 

We assume that GRBs trace the star formation history and we adopt, as our ``standard'' model
for $z\la 10$, the Rowan-Robinson star formation rate
which can be fitted with the expression
\begin{equation}
 R_{\rm GRB}(z)  = \left\{
  \begin{array}{ll}
     R_0\; 10^{0.75z}, & \hbox{$ z < z_{\rm peak}$} \\
   R_0 \;10^{0.75 z_{\rm peak}},   & \hbox{ $z \ge z_{\rm peak}$}   \\
  \end{array}\right.\;,
\end{equation}
where $z_{\rm peak}\sim 2$. For $z\ga 10$, in our standard model we use an interpolation
that follows the star formation history derived in numerical simulations by Gnedin \& Ostriker (1997).
While using this SFR as our working model, we also explore the effects on the predicted 
distribution $dn/d\theta$ of different star formation histories, and in particular 
we consider two opposite extremes, one in which the SFR does not rapidly decline for  
$z\ga 10$ as implied in the numerical simulations of Gnedin \& Ostriker, and another, 
the Madau curve (Madau 1996), in which the SFR rapidly declines at redshifts $z\ga 3$. 

For each model for the SFR, the normalization constant $R_0$ is determined
by the condition
\beq
\int_0^{\pi/2}d\theta \;\frac{dn(\theta)}{d\theta}\;=\;R_{\rm GRB}^{\rm obs}\;,
\eeq
where 
$R_{\rm GRB}^{\rm obs}=667$ yr$^{-1}$ is the observed BATSE rate, and
$F_{\rm lim}$ in Eq. (\ref{eq:flim}) is the BATSE threshold flux for which
this rate has been measured.  We adopted the 90\% efficiency
peak flux threshold for BATSE, that is $F_{\rm lim}=0.424$ ph/sec (e.g.
Mallozzi, Pendleton \& Paciesas 1996).

\section{Comparison with data and predictions for more sensitive surveys}

In order to compare the theoretical distributions derived in \S{2.2},
we require a sample of gamma-ray bursts whose values of $\theta_{\rm
break}$ have been measured. The largest published sample of
$\theta_{\rm break}$ values at the time of our work comes from the
analysis of 28 bursts with redshifts and well-studied afterglow light
curves by Bloom et al. (2003). This list of 28 includes {\it all}
bursts with measured redshifts at this time. For all but four, some
limit on $\theta_{\rm break}$ was derived. The exceptions are
GRB~970228, in which sparse data together with likely contribution
from a supernovae makes it difficult to interpret the lightcurve,
GRB~990506 where no sufficient data exists, GRBs~980425 which had no
optical counterpart, and GRB~021211 which is being analyzed at the
time of writing.  For eight of the remaining 24, upper or lower limits
where put on the opening angle. We have not used these limits in our
comparison. One could think that this would tend to have the effect of
narrowing our sample distribution. However, the upper and lower limits
do not tend to be at the edge of the distribution of the measured
opening angles (see Table 2 of Bloom et al. 2003). They therefore do
not necessarily reflect extreme cases, but cases with lower quality
data. However, it should be noted that in three cases (GRBs~970828,
991216, 990705) the confidence that indeed a jet has been identified
is weak, since the break was observed in a single frequency only.
Finally, we should remark that the inferred values of $\theta_{\rm
break}$ have some uncertainties. These values are indeed computed
using the expression given in Frail et al. (2001): $\theta_{\rm break}
\propto t_{\rm break}^{3/8} (1+z)^{-3/8}E_{\rm
iso}^{-1/8}\eta_\gamma^{1/8}n^{1/8}$. The measured values of $t_{\rm
break}$ have relative errors within 30\%, while for the densities a
value of $n=10$ cm$^{-3}$ is assumed for the 5 (out of 16) cases for
which the data quality did not allow a self-consistent determination
of the density through broad-band afterglow modelling. Finally, Bloom
et al. assumed a constant value ($\eta_\gamma=0.2$) for the efficiency
of the bursts. Whereas the dependence on $\eta_\gamma$ is weak, a large
spread in this not-well constrained quantity would introduce a further
source of error in the determination of $\theta_{\rm break}$.  The
combination of these caveats prevents a solid comparison of the data
with our predictions, and the following comparison should be taken as
a general guide for this type of analysis, while showing that the
observed distribution so far does not seem to be in contradiction with
the prediction of the structured jet hypothesis.

We performed a Kolmogorov-Smirnov (KS) test to assess the
compatibility of the theoretical distributions with the data, and we
found (see Figure 1), for the Rowan-Robinson SFR, a probability of
$\sim 90\%$ that the data are drawn from the theoretical
distribution. In order to have a reasonable agreement with the data, a
value $T\sim 8$ sec is needed in the theoretical model.

As shown in Figure 1, there is little difference in the results (and
in the required value of $T$ to fit the data) between the case where
the SFR drops rapidly after $z\ga 10$ (model 1), and where it keeps
constant also at higher redshifts (model 2). The similarity between
the distributions in model 1 and model 2 is a consequence of the
combined effect of the decrease in volume at those high redshifts and
the increased time dilution of the observed rate.  On the other hand,
for the same value of $T$ (or equivalently peak luminosity), the
distribution that uses the Madau SFR (model 3) predicts significantly
more events with large angles. This is because this star formation
rate drops abruptly at a redshift $z\ga 3-4$, and therefore there is
no much gain in the number of small-$\theta$ (i.e. brighter) events
which can be seen at higher redshifts. A KS test showed that the model
3 distribution is consistent with the data (at the 40\% level) if a
value $T\sim 25$ is used. A lower value for the peak luminosity is
needed to shift the $dn/d\theta$ distribution to lower $\theta$'s.

For a jet model to be self-consistent it is necessary that, if the
distribution $dn/d\theta$ has a good agreement with the data for
certain model parameters, the corresponding $dn/dS$ has to have a good
agreement with the corresponding data for the same set of parameters.
We found that with $T\sim 8$ (as required in model 1) the peak fluxes
are within a factor of a few for those bursts with measured
$\theta$. More generally, when comparing the theoretical $dn/dS$
distribution with the all BATSE catalogue we find a very good
agreement for peak fluxes in the range $1\la P\la 15$ ph/cm$^2$/s, and
a departure (as an overprediction) at higher fluxes. However, it
should be noted that, given the large number of bursts in the BATSE
catalogue, a comparison with the $dn/dS$ data distribution is much
more sensitive to the model parameters than the $dn/d\theta$
comparison. In a recent analysis, Lloyd et al. (2001) found that a
good fit can be obtained with a luminosity function $\propto L^{-2.2}$
(which would reduce the number of high $P$ bursts with respect to our
model) and a redshift evolution. Such a detailed analysis is beyond
the scope of this paper given the lack of comparable wealth of data
for the $\theta$ distribution
\footnote{It should also be noted that the value of the peak
luminosity that best fits the $dn/d\theta$ distribution depends on the
adopted value of the detection efficiency, which may vary from
burst to burst.}.  On the other hand, we should remark that, if we
adopt the value $T\sim 25$ required to fit the $dn/d\theta$
distribution with the Madau SFR, then the luminosities that we predict
are generally smaller than those measured for
the bursts with known redshift. Therefore, in the following we will
only use our model 1 with $T= 8$ for further calculations.  The
corresponding cumulative distribution is shown with the solid line in
the two panels in Figure 2 where it is compared to the binned,
cumulative data. We should note that the size of the bins
in the figure has been chosen so that there is an equal increment
in the distribution for every new data point. This gives a better
visual idea of the data distribution, but does not reflect the
actual magnitude of the errors as described above.
     
Figure 3 shows the predicted evolution of the distribution of observed
break angles with increasing sensitivity of the survey. In particular,
we considered the 100\% efficiency sensitivities corresponding to
BATSE (solid line), HETE-2 and {\em Swift} (dashed line), and an
intermediate sensitivity (dotted line). An interesting prediction of
the luminosity distribution in Eq.(\ref{eq:Lp0}) is that the average
observed jet angle is an increasing function of the survey
sensitivity. There are two counteracting effects that determine the
average observed jet angle $<\theta>$ of the sample as a function of
the survey sensitivity. As $F_{\rm lim}$ decreases, $z_{\rm max}$
increases, bringing into the sample a fraction of bursts with larger
redshifts and correspondingly smaller $\theta$ (which are the most
luminous).  On the other hand, the higher sensitivity also brings into
the sample a fraction of bursts with larger $\theta$ at lower
redshifts, and this latter effect dominates over the former, partly as
a result of the volume-reduced and time-dilated rate of the high-$z$
bursts.

\section{Extensions}

All the above results have been produced under the assumption of a
strict correlation between the total energy of the burst and its peak
flux.  However, as discussed in \S 2, this relation has a scatter;
therefore we have also investigated the extent to which our results
change when a dispersion in the distribution of the values of $T$
(i.e. in the relation ~\ref{eq:Lp0}) is introduced.  If we
parameterize the scatter in $T$ with a probability distribution,
$P(T)$, then the distribution of jet angles (\ref{eq:dndt}) is
generalized to
\begin{equation}
\frac{dn(\theta)}{d\theta} = 2\pi\sin\theta 
\int_0^\infty dT\; P(T)
\int_0^{z_{\rm max}(\theta,T)}
dz\; \frac{R_{\rm GRB}(z)}{(1+z)}\; \frac{dV(z)}{dz}\;.
\label{eq:dndt1}
\end{equation}
We took the probability distribution for the scatter, $P(T)$, to be
a log-gaussian distribution with mean equal to $T=8$, and studied
the dependence of the break-angle distribution $dn/d\theta$ on the
width of the distribution $\sigma_{T}$.  
A K-S test shows that, with a scatter $\sigma_{T}=0.3$, the theoretical 
distribution is still compatible with the data at the 80\%, while 
with $\sigma_{T}=0.5$ the agreement is at the 
40\% level.

All the results so far have been derived under the assumption of an
energy distribution from the jet axis $\propto\theta^{-2}$ in the
interval $0\le\theta\le \pi/2$. However, close to the axis this
divergence must naturally have a cutoff which we represent by a core
of size $\theta_c$. We now explore how our results vary by allowing
for the presence of a core in the inner part of the jet and an outer
cutoff at some large angle $\theta_j$, where the luminosity drops
rapidly to zero rather than following the profile in
Eq.~(\ref{eq:Lp0}).  In this case the peak photon luminosity is given
by $L_P(\theta)$ as in Eq. (\ref{eq:Lp0}) for
$\theta_c<\theta<\theta_j$, and by $L_P(\theta=\theta_c)$ for all the
angles $\theta\le\theta_c$. Hence the cumulative $\theta$ distribution
is given by
\begin{equation}
 N(<\theta)  = \left\{
  \begin{array}{ll}
     0 & \hbox{$ \theta < \theta_c$} \\
   \int_0^{\theta_c}d\theta'\;\frac{dn[\theta',L(\theta_c)]}{d\theta'}\equiv N_c   
& \hbox{ $\theta = \theta_c$}   \\
N_c \,+\, \int_{\theta_c}^{\theta}d\theta'\;\frac{dn[\theta',L(\theta')]}{d\theta'}\equiv N_j
& \hbox{ $\theta_c< \theta < \theta_j$}  \\
  N_j & \hbox{$ \theta > \theta_j$} \\
  \end{array}\right.\;.
\end{equation}

For the case where the increase in luminosity saturates at an angle
$\theta_c$ from the jet axis (left panel of Fig. 2), the number of
observed bursts with break angle $\theta<\theta_c$ is smaller than the
corresponding number at the same observed $\theta$ for the
distribution with $\theta_c=0$ (solid line in the figure). This is
because $z_{\rm max}$ saturates to the value given by the solution of
Eqn.~(\ref{eq:flim}) with $L=L_c$ for all $\theta\le\theta_c$, whereas
when there is no core\footnote{Note that when we say here ``no core''
or $\theta_c=0$, we mean an infinitesimally small core, as there would be
a formal divergence in the energy if $\theta_c$ were precisely equal to
zero.}, $z_{\rm max}$ is larger for the jet angles $\theta<\theta_c$.
The situation is reversed in the case of a jet with $\theta_c=0$ but a
total aperture $\theta_j<\pi/2$ (right panel of Fig.2).  The number of
bursts with observed break angle $\theta>\theta_j$ is zero.
Therefore, in order for the normalization (i.e. total number of
observed bursts) to be the same for any $\theta_j$, the cumulative
number $N(<\theta)$ for the case $\theta_j<\pi/2$ must be larger than
the corresponding number for the case $\theta_j=\pi/2$ (solid line in
the figure).

The probability remains of the same order by varying the core angle
$\theta_c$ in the range $0\la\theta_c\la 0.055$, and drops as
$\theta_c$ is increased above 0.055 (which is the smallest angle in
the data set).  Similarly, no significant variation in the probability is
found as the outer boundary of the jet angle, $\theta_j$ is decreased
from $\pi/2$ to 0.55 rad, which is the largest observed break angle.
In short, the current data is too poor to constrain either
the size of the core, $\theta_c$, or the outer size of the jet,
$\theta_j$, beyond the trivial statement that this range must include
the range of observed opening angles. It should however be remarked
that, whereas the current data on the observed $\theta_{\rm break}$ do not allow
to pin down the values of the model parameters $\theta_c$ and
$\theta_j$, the type of analysis that we are proposing has the
potentiality to further constrain details of the model once a larger
sample of jet opening angles is gathered.

\section{Conclusions}

A natural framework for the interpretation of achromatic breaks in the
afterglow light curves is the presence of jets in the GRB
ejecta. There is however a certain degree of degeneracy in the
resulting light curves between the case where the jet has uniform
energy profile and a sharp edge, and where it has instead a variable
energy profile. Distinguishing between the two scenarios is especially
important in order to have a proper estimate of the GRB event rates,
and a better understanding of the physics of the GRB explosion.  A
particularly useful discriminant of the two jet scenarios is the
distribution of the observed break times, which can be theoretically
predicted in the structured jet model, and compared to the data.  

In this paper, we have derived such distribution and compared it to
the observed sample of data on $\theta_{\rm break}$.  We found that
the observed data set, altough a small sample, is consistent with the
predictions of the structured-jet model.  We should however remark
that the alternative ``uniform'' model, in which there is a direct
correlation between the observed opening angle and the physical jet
angle, cannot be ruled out by this approach. However, if, as more data
becomes available, the agreement with the predictions of the
structured jet model remains intact, it would be very contrived to
justify it within the framework of the uniform jet. This model, in fact,
does not make any prediction for the distribution of
opening angles.  One would then need to invoke an explanation for why
there is the same number of bursts for each logarithmic
interval of openening angle.

Besides performing a first attempt to test the structured-jet model,
we have shown that this model offers a number of
predictions which may be testable with future, larger datasets. The
predicted opening angle distribution shown in Figure 1 (which assumes
the BATSE detection threshold), has a distinct peak at
$\theta\sim0.12$ rad. We predict that future, more sensitive missions will
predict more bursts with large opening angles. Specifically, we
estimate that the opening angle distribution for the SWIFT mission
will peak near $\theta\sim0.3$ rad. Surprisingly, in this model the
average redshift is only a weak function of the sensitivity and
consequently we expect no increase in the {\it average} redshift
detected by the next generation of gamma-ray instruments.

Another prediction that the structured-jet model makes, regards the
number of GRBs in the local universe.  Our ``standard'' SFR model
yields $R_{\rm GRB}(z=0)\sim 0.5$ Gpc$^{-3}$ yr$^{-1}$. Combining this
with the local galactic density $\approx 0.0048$ Mpc$^{-3}$ (Loveday
et al. 1992) one obtains $\approx 0.1$ GEM (galactic events per
Myr). This is an interesting number for detection of local GRB
remnants (Loeb \& Perna 1998; Efremov et al. 1998; Perna et
al. 2000). For the uniform jet model, precise rates are more difficult
to estimate, as they depend on the assumed luminosity function, or
equivalently the intrinsic jet angle distribution which is not known
apriori for this model. If one were to assume that $dn/d\theta
\propto\theta$ as in the structured jet model, this would result in a
logarithmic correction factor, $(1+\log[(\pi/2)/\theta_c])$, to the
rate that we estimated here for the structured jet model\footnote{The
correction is expected to be less than an order of magnitude as
$\theta_c\ga 1/\Gamma$.}. The total energy output of GRBs would
however remain the same, as in the structured jet model the total
energy of each burst is also corrected by a logarithmic factor with
respect to the total energy of each burst in the uniform case.

\acknowledgements We thank the referee for a very careful and
thoughtful review of our manuscript.  RP and DAF thank the California
Institute of Technology for its kind hospitality during the time that
part of this work was carried out.  RS holds a senior Sherman
Fairchild fellowship. This research was partially supported by a NASA
grant to RS.  The NRAO is a facility of the National Science
Foundation operated under cooperative agreement by Associated
Universities, Inc.

\newpage

\begin{figure}[t]
\plotone{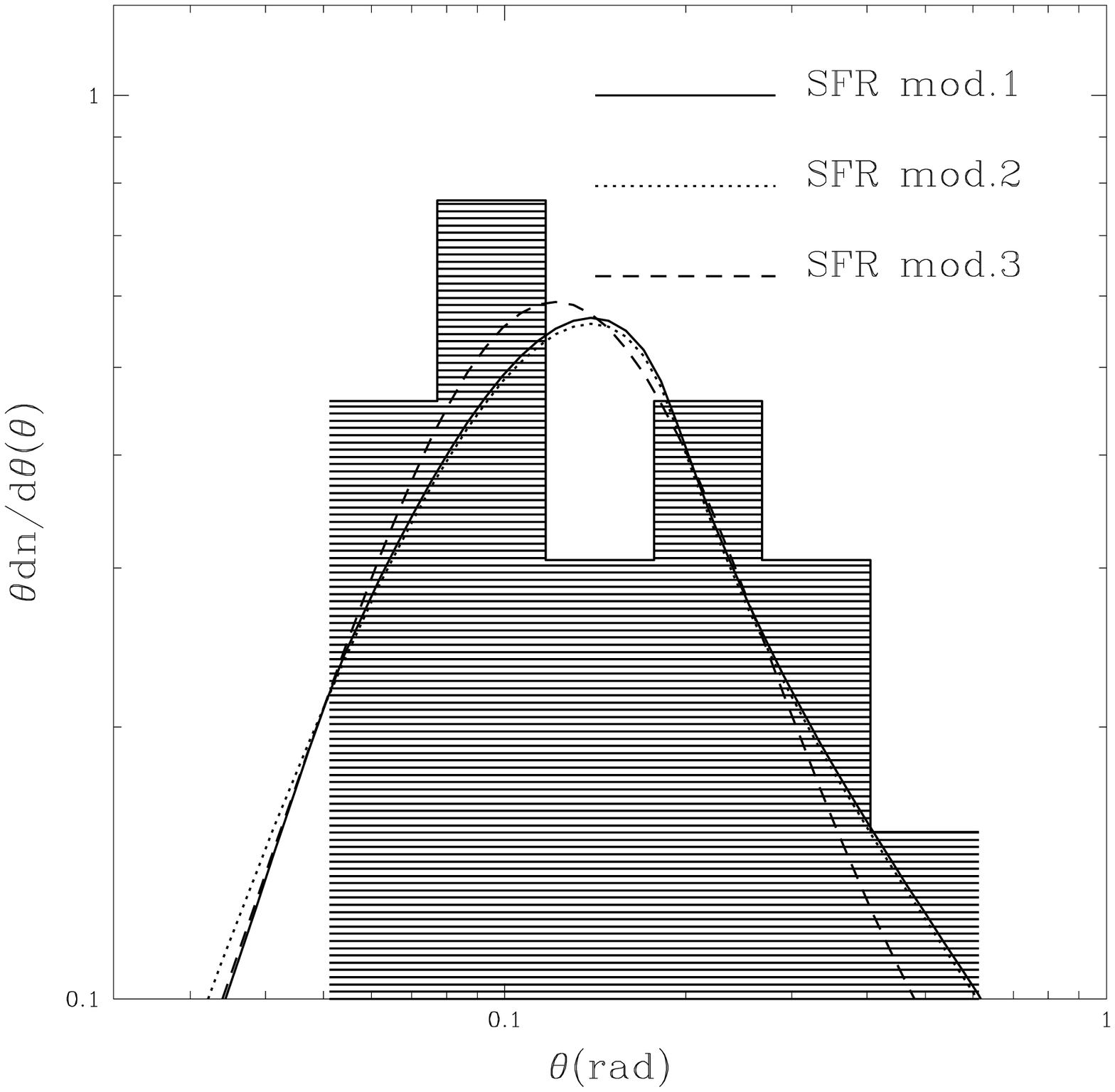}
\caption{Distribution of observed jet angles for different star
formation rates: in model 1 the Rowan-Robinson SFR is assumed up to
$z\ge 10$, and a rapid drop is assumed at larger redshifts as in the
numerical simulations of Gnedin \& Ostriker. In model 2, no dropout is
assumed for $z\ga 10$, while model 3 uses the Madau SRF. The histogram
shows the observed distribution from the data available so far for a
sample of 16 bursts.}

\end{figure} 

\begin{figure}[t]
\plottwo{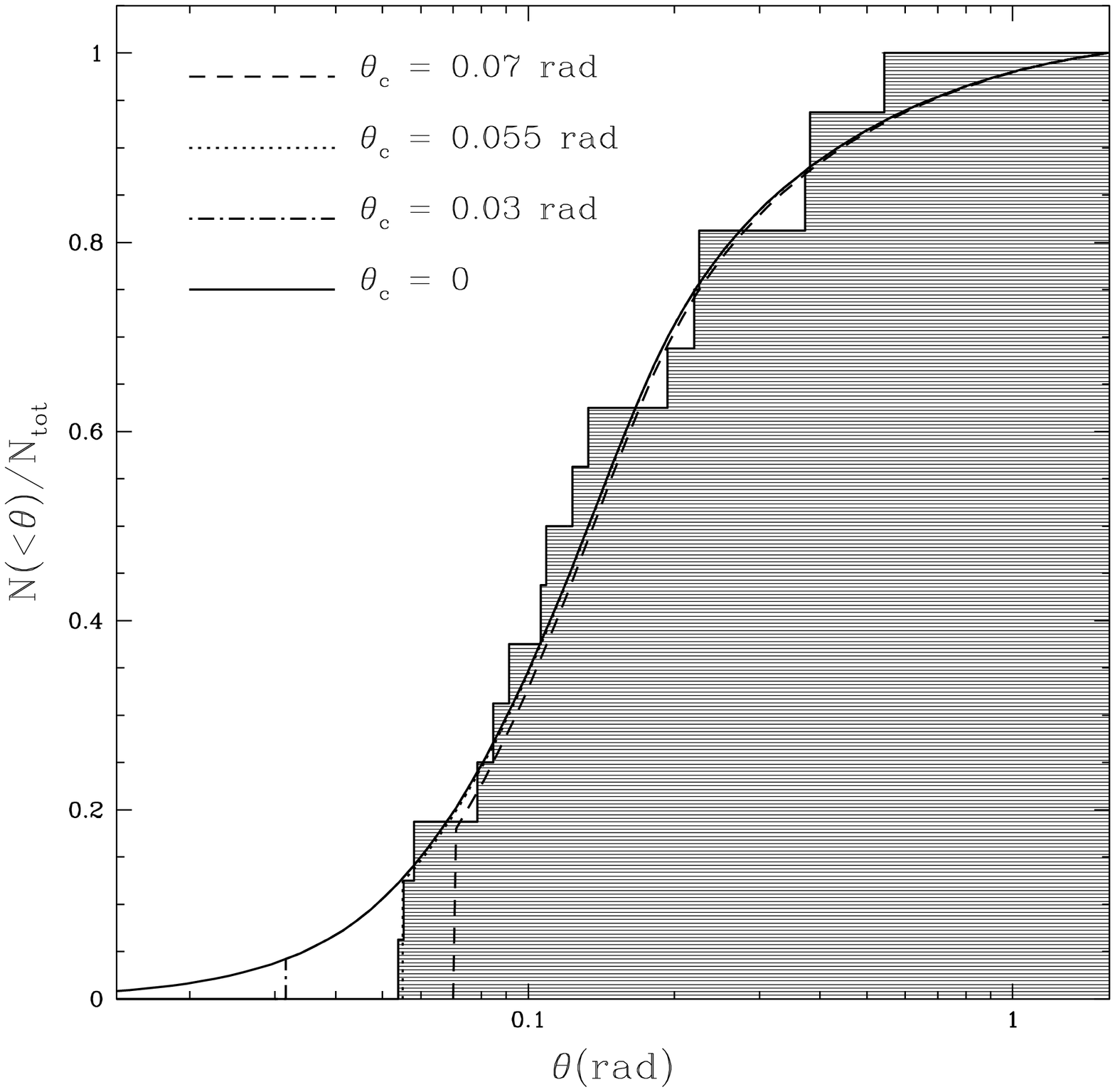}{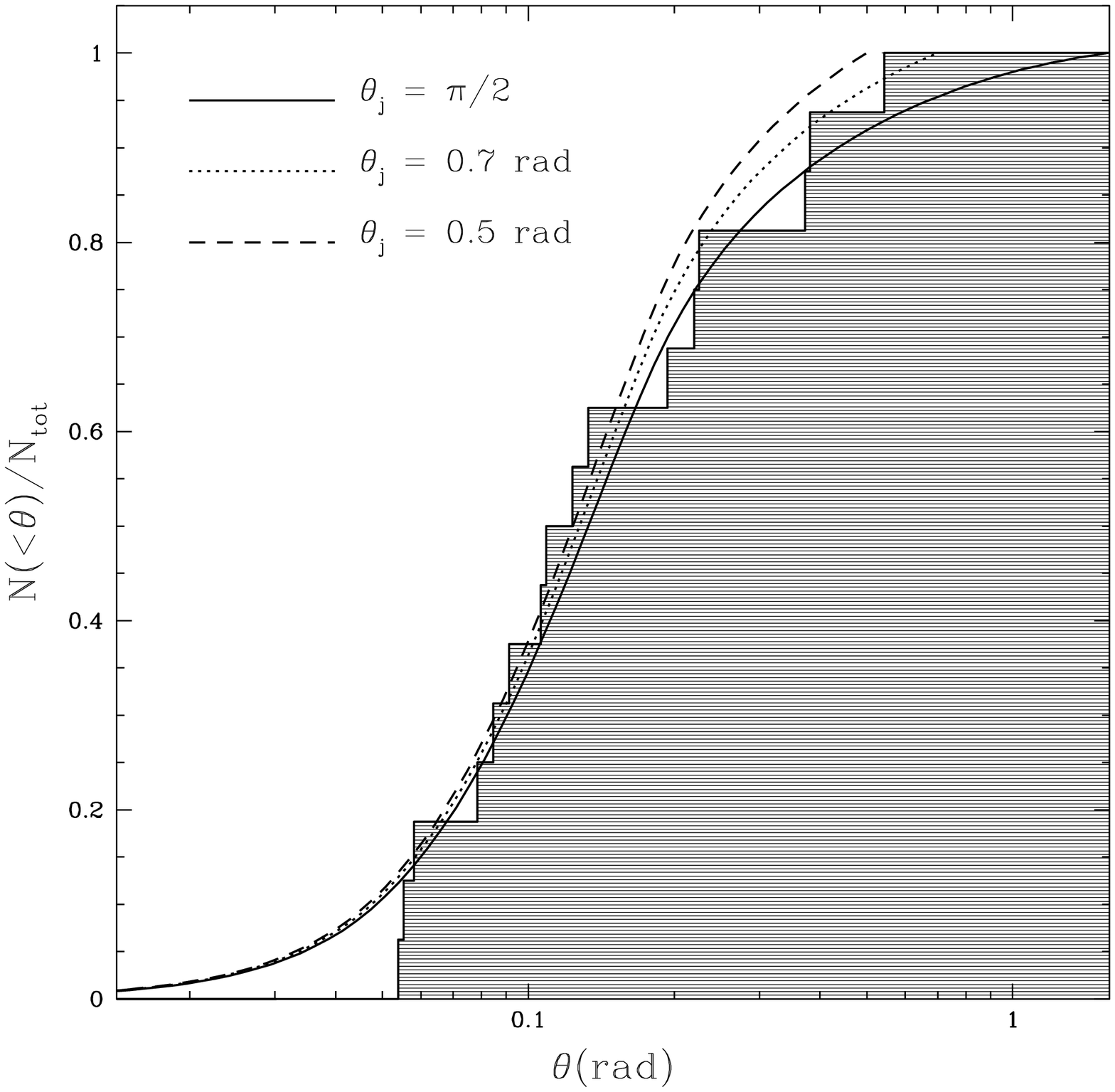}
\caption{Left panel: Cumulative distribution for the SFR model 1, and
various values of the jet core. Here $\theta_j=\pi/2$. Right panel:
Cumulative distribution for the SFR model 1, and various values of the
jet aperture. Here $\theta_c=0$.}
\end{figure} 

\begin{figure}[t]
\plotone{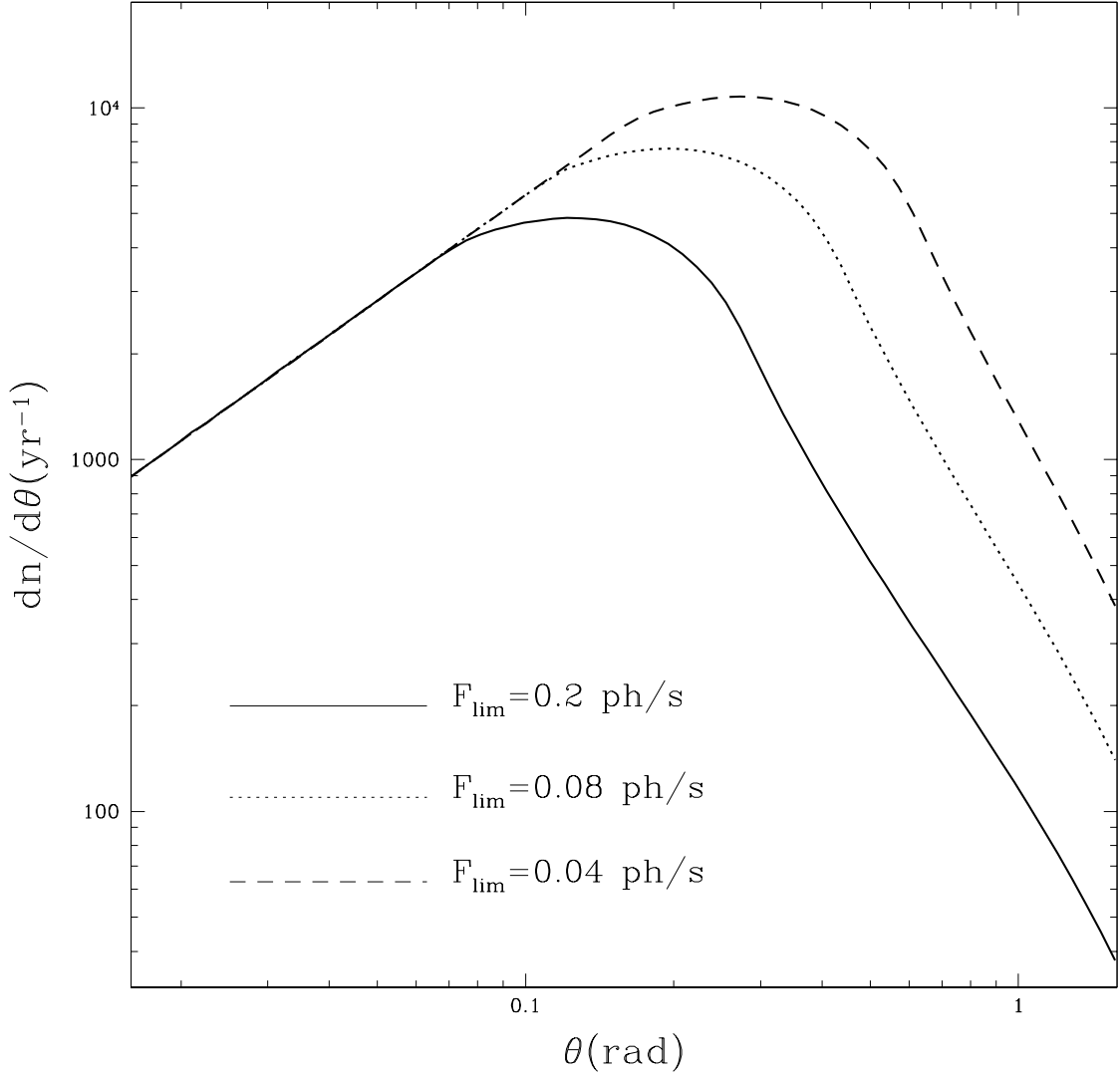}
\caption{Probability distribution for the observed jet angle 
$\theta$ for different values of the survey sensitivity threshold.
The higher the sensivity of the survey, the larger is the mean beaming
angle $\theta$ that is observed. Here $\theta_j=\pi/2$ and $\theta_c=0$
for the SFR model 1.}  

\end{figure}

\end{document}